# On spherically symmetric metric satisfying the positive kinetic energy coordinate condition

T. Mei

(Department of Journal, Central China Normal University, Wuhan, Hubei PRO, People's Republic of China
E-Mail: meitao@mail.ccnu.edu.cn meitaowh@public.wh.hb.cn )

**Abstract** Generally speaking, there is a negative kinetic energy term in the Lagrangian of the Einstein-Hilbert action of general relativity; On the other hand, the negative kinetic energy term can be vanished by designating a special coordinate system. For general spherically symmetric metric, the question that seeking special coordinate system that satisfies the positive kinetic energy coordinate condition is referred to solving a linear first-order partial differential equation. And then, we present a metric corresponding to the Reissner-Nordström solution that satisfies the positive kinetic energy coordinate condition. Finally, we discuss simply the case of the Tolman metric.

At first, we cite two conclusions in Ref. [1]:

① For an arbitrary metric indicated by the line element

$$\mathrm{d}s^2 = g_{\mu\nu}\mathrm{d}x^\mu \mathrm{d}x^\nu , \tag{0-1}$$

the corresponding negative kinetic energy term of gravitation field is

$$L_{\mathrm{GNK}} = \frac{2}{3g}\left[\left(\sqrt{|g_{ij}|}\frac{g^{0\lambda}}{g^{00}}\right)_{,\lambda}\right]^2 , \tag{0-2}$$

where $g = |g_{\mu\nu}| < 0$, $|g_{ij}|$ is the determinant of the 3- dimensional metric $g_{ij}$.

For the Schwarzschild metric indicated by the line element

$$\mathrm{d}s^2 = -\left(1-\frac{r_s}{r}\right)\mathrm{d}t^2 + \frac{1}{1-\frac{r_s}{r}}\mathrm{d}r^2 + r^2\left(\mathrm{d}\theta^2 + \sin^2\theta\mathrm{d}\varphi^2\right), \tag{0-3}$$

where $r_s = 2m$, we see that $\frac{\partial |g_{ij}|}{\partial t} = \frac{\partial}{\partial t}\left(\frac{r^4\sin^2\theta}{1-\frac{r_s}{r}}\right) = 0$ in the area of $r_s < r$, hence, according to (0-2), in the area of $r_s < r$ the negative kinetic energy term of gravitation field vanishes.

But (0-3) cannot be continued into the area of $0 < r < r_s$. For continuing (0-3) into the area of $0 < r < r_s$, one has used the method of coordinate transformation and obtained some metrics, e.g., the Lemaitre[2] and the Kruskal[3] metrics. However, for the Lemaitre and the Kruskal metrics of the Schwarzschild solution, using (0-2), we can verify easily that there are corresponding negative kinetic energy terms of gravitation field in total space, respectively.

For the Robertson-Walker metric[4] indicated by the line element

$$ds^2 = -dt^2 + R^2(t)\left[\frac{dr^2}{1-kr^2} + r^2\left(d\theta^2 + \sin^2\theta d\varphi^2\right)\right], \tag{0-4}$$

we have $g^{00} = -1$, $g^{0i} = 0$, $\left|g_{ij}\right| = R^6(t)\frac{r^4\sin^2\theta}{1-kr^2}$, according to (0-2), the corresponding negative kinetic energy term of gravitation field in total space is $-\frac{6}{R^2(t)}\left(\frac{dR(t)}{dt}\right)^2$.

② One of the positive kinetic energy coordinate conditions is

$$\left(\sqrt{\left|g_{ij}\right|}\frac{g^{0\lambda}}{g^{00}}\right)_{,\lambda} = 0. \tag{0-5}$$

In this paper, at first, we discuss some characteristics of general spherically symmetric metric that satisfy the positive kinetic energy coordinate condition given by (0-5). And then, we present a metric corresponding to the Reissner-Nordström solution that satisfies (0-5).

## 1 General spherically symmetric metric

For general case of spherically symmetric metric indicated by the line element[4]

$$ds^2 = -K^2(t,r)dt^2 + Q^2(t,r)dr^2 + R^2(t,r)(d\theta^2 + \sin^2\theta d\varphi^2), \tag{1-1}$$

$\left|g_{ij}\right| = Q^2(t)R^4\sin^2\theta$, and the positive kinetic energy coordinate condition (0-5) is

$$\left(\sqrt{\left|g_{ij}\right|}\frac{g^{0\lambda}}{g^{00}}\right)_{,\lambda} = \sin\theta\frac{\partial\left(QR^2\right)}{\partial t} = 0.$$

If this condition is not satisfied, then we can consider the following coordinate transformation

$$t = t(\rho,\sigma),\ r = r(\rho,\sigma), \tag{1-2}$$

under the transformation (1-2), (1-1) becomes

$$\begin{aligned} ds^2 = &-W^2(\rho,\sigma)d\sigma^2 + 2E(\rho,\sigma)d\sigma d\rho + \Omega^2(\rho,\sigma)d\rho^2 \\ &+ R^2(t(\rho,\sigma))(d\theta^2 + \sin^2\theta d\varphi^2), \end{aligned} \tag{1-3}$$

where,

$$W^2(\rho,\sigma) = K^2(t'_\sigma)^2 - Q^2(r'_\sigma)^2, \tag{1-4}$$

$$\Omega^2(\rho,\sigma) = -K^2(t'_\rho)^2 + Q^2(r'_\rho)^2, \tag{1-5}$$

$$E(\rho,\sigma) = -K^2 t'_\rho t'_\sigma + Q^2 r'_\rho r'_\sigma. \tag{1-6}$$

(1-4)×(1-5) +(1-6)$^2$, we have

$$W^2\Omega^2 + E^2 = K^2Q^2(r'_\rho t'_\sigma - t'_\rho r'_\sigma)^2. \tag{1-7}$$

The corresponding metric indicated by (1-3) are

$$g_{00} = -W^2, g_{01} = g_{10} = E, g_{11} = \Omega^2, g_{22} = R^2, g_{33} = R^2\sin^2\theta; g_{\mu\nu} = 0, \text{others}. \tag{1-8}$$

$$\begin{aligned} &g^{00} = -\frac{\Omega^2}{W^2\Omega^2 + E^2}, g^{01} = g^{10} = \frac{E}{W^2\Omega^2 + E^2}, g^{11} = \frac{W^2}{W^2\Omega^2 + E^2}, \\ &g^{22} = \frac{1}{R^2}, g^{33} = \frac{1}{R^2\sin^2\theta};\ g^{\mu\nu} = 0, \text{others}; \end{aligned} \tag{1-9}$$

$\sqrt{\left|g_{ij}\right|} = \Omega R^2\sin\theta$ and the positive kinetic energy coordinate condition (0-5) becomes

$$\left(\sqrt{|g_{ij}|}\frac{g^{0\lambda}}{g^{00}}\right)_{,\lambda}=\sin\theta\left[\left(\Omega R^2\right)'_\sigma-\left(\frac{ER^2}{\Omega}\right)'_\rho\right]$$

$$=\sin\theta\frac{R}{\Omega^3}\left\{\Omega^2\left[\frac{1}{2}R\left(\Omega^2\right)'_\sigma+2\Omega^2R'_\sigma-RE'_\rho-2ER'_\rho\right]+\frac{1}{2}RE\left(\Omega^2\right)'_\rho\right\}=0;$$

Substituting $\Omega^2$ and $E$ given by (1-5) and (1-6) respectively to the above formula, and notice that

$$\begin{aligned}&K'_\sigma=\dot{K}t'_\sigma+K'r'_\sigma\,,\ K'_\rho=\dot{K}t'_\rho+K'r'_\rho;\\&Q'_\sigma=\dot{Q}t'_\sigma+Q'r'_\sigma\,,\ Q'_\rho=\dot{Q}t'_\rho+Q'r'_\rho\,;\\&R'_\sigma=\dot{R}t'_\sigma+R'r'_\sigma\,,\ R'_\rho=\dot{R}t'_\rho+R'r'_\rho\,;\end{aligned}\tag{1-10}$$

where $\dot{K}=\frac{\partial K(t,r)}{\partial t}$, $K'=\frac{\partial K(t,r)}{\partial r}$, etc; we have

$$\left(\sqrt{|g_{ij}|}\frac{g^{0\lambda}}{g^{00}}\right)_{,\lambda}=\sin\theta\frac{K^2Q^2R^2}{\Omega^3}\left(r'_\rho t'_\sigma-t'_\rho r'_\sigma\right)$$

$$\times\left[\frac{Q^2}{K^2}\left(\frac{\dot{Q}}{Q}+2\frac{\dot{R}}{R}\right)\left(r'_\rho\right)^3+\left(2\frac{K'}{K}-\frac{Q'}{Q}+2\frac{R'}{R}\right)t'_\rho\left(r'_\rho\right)^2\right.$$

$$\left.+\left(\frac{\dot{K}}{K}-2\frac{\dot{Q}}{Q}-2\frac{\dot{R}}{R}\right)\left(t'_\rho\right)^2r'_\rho-\frac{K^2}{Q^2}\left(\frac{K'}{K}+2\frac{R'}{R}\right)\left(t'_\rho\right)^3+\left(r'_\rho t''_{\rho\rho}-t'_\rho r''_{\rho\rho}\right)\right]=0,$$

from (1-7) we know that $r'_\rho t'_\sigma-t'_\rho r'_\sigma\neq0$, and because $r'_\rho t''_{\rho\rho}-t'_\rho r''_{\rho\rho}=\left(r'_\rho\right)^2\left(\frac{t'_\rho}{r'_\rho}\right)'_\rho$, we therefore obtain the positive kinetic energy condition for (1-3):

$$\begin{aligned}&\frac{Q^2}{K^2}\left(\frac{\dot{Q}}{Q}+2\frac{\dot{R}}{R}\right)\left(r'_\rho\right)^3+\left(2\frac{K'}{K}-\frac{Q'}{Q}+2\frac{R'}{R}\right)t'_\rho\left(r'_\rho\right)^2\\&+\left(\frac{\dot{K}}{K}-2\frac{\dot{Q}}{Q}-2\frac{\dot{R}}{R}\right)\left(t'_\rho\right)^2r'_\rho-\frac{K^2}{Q^2}\left(\frac{K'}{K}+2\frac{R'}{R}\right)\left(t'_\rho\right)^3+\left(r'_\rho\right)^2\left(\frac{t'_\rho}{r'_\rho}\right)'_\rho=0.\end{aligned}\tag{1-11}$$

The above form enlightens us that we can try to take

$$t'_\rho=F(t,r)r'_\rho,\tag{1-12}$$

and notice that

$$\left(\frac{t'_\rho}{r'_\rho}\right)'_\rho=\frac{\partial F(t,r)}{\partial\rho}=\frac{\partial F(t,r)}{\partial r}r'_\rho+\frac{\partial F(t,r)}{\partial t}t'_\rho=\frac{\partial F(t,r)}{\partial r}r'_\rho+\frac{\partial F(t,r)}{\partial t}F(t,r)r'_\rho,$$

substituting the above formula and (1-12) to (1-11), we have

$$\begin{aligned}&\frac{Q^2}{K^2}\left(\frac{\dot{Q}}{Q}+2\frac{\dot{R}}{R}\right)+\left(2\frac{K'}{K}-\frac{Q'}{Q}+2\frac{R'}{R}\right)F(t,r)+\left(\frac{\dot{K}}{K}-2\frac{\dot{Q}}{Q}-2\frac{\dot{R}}{R}\right)F^2(t,r)\\&-\frac{K^2}{Q^2}\left(\frac{K'}{K}+2\frac{R'}{R}\right)F^3(t,r)+\frac{\partial F(t,r)}{\partial r}+F(t,r)\frac{\partial F(t,r)}{\partial t}=0.\end{aligned}\tag{1-13}$$

Once we obtain a solution $F(t,r)$ of the linear first-order partial differential equation (1-13), according to (1-12) we obtain an equation on $t(\rho,\sigma)$ and $r(\rho,\sigma)$:

$$\frac{\partial t(\rho,\sigma)}{\partial\rho}=F(t(\rho,\sigma),r(\rho,\sigma))\frac{\partial r(\rho,\sigma)}{\partial\rho}.\tag{1-14}$$

And, further, we can try to obtain the forms of $t(\rho,\sigma)$ and $r(\rho,\sigma)$ based on (1-14).

We can verify easily that

$$F(t,r)=\frac{Q(t,r)}{K(t,r)} \tag{1-15}$$

is a solution of (1-13). However, if we adopt (1-15), then according to (1-12) and (1-5) we have $\Omega(\rho,\sigma)=0$ . Hence, (1-15) is not applicable.

However, (1-15) enlightens us that we can take

$$F(t,r)=\frac{Q(t,r)}{K(t,r)}J(t,r)\,; \tag{1-16}$$

According to (1-12) and (1-16), (1-5) become

$$\Omega^2(\rho,\sigma)=Q^2\left(1-J^2\right)(r'_\rho)^2\,, \tag{1-17}$$

from the above expression we see that $J(t,r)$ must satisfy

$$J^2(t,r)<1\,. \tag{1-18}$$

Substituting (1-16) to (1-13), we obtain

$$\frac{Q}{K}\left(\frac{\dot{Q}}{Q}+2\frac{\dot{R}}{R}\right)\left(1-J^2\right)+\left(\frac{K'}{K}+2\frac{R'}{R}\right)J\left(1-J^2\right)+\frac{Q}{K}J\frac{\partial J}{\partial t}+\frac{\partial J}{\partial r}=0\,.$$

We must seek a solution $J(t,r)$ ( $J(t,r)\neq 1$ ) of the above equation after the concrete forms of $K(t,r)$, $Q(t,r)$, $R(t,r)$ are given. And, considering (1-18), the above equation can be written to the following form:

$$\frac{\partial}{\partial r}\left(\frac{KR^2J}{\sqrt{1-J^2}}\right)+\frac{\partial}{\partial t}\left(\frac{QR^2}{\sqrt{1-J^2}}\right)=0\,. \tag{1-19}$$

We therefore can set

$$J(t,r)=\frac{\Theta(t,r)}{\sqrt{\Theta^2(t,r)+K^2(t,r)R^4(t,r)}}\,, \tag{1-20}$$

it is obvious that the form of $J(t,r)$ given by (1-20) satisfies (1-18). Substituting (1-20) to (1-19), we obtain an equation that $\Theta(t,r)$ satisfies:

$$\frac{\partial\Theta}{\partial r}+\frac{\partial}{\partial t}\left(\frac{Q}{K}\sqrt{\Theta^2+K^2R^4}\right)=0\,. \tag{1-21}$$

Once we obtain $\Theta(t,r)$ by (1-21), (1-14) and (1-17) become

$$\begin{aligned}\frac{\partial t(\rho,\sigma)}{\partial\rho}&=\frac{Q(t(\rho,\sigma),r(\rho,\sigma))}{K(t(\rho,\sigma),r(\rho,\sigma))}\\&\times\frac{\Theta(t(\rho,\sigma),r(\rho,\sigma))}{\sqrt{\Theta^2(t(\rho,\sigma),r(\rho,\sigma))+K^2(t(\rho,\sigma),r(\rho,\sigma))R^4(t(\rho,\sigma),r(\rho,\sigma))}}\frac{\partial r(\rho,\sigma)}{\partial\rho}\,,\end{aligned} \tag{1-22}$$

$$\Omega^2(\rho,\sigma)=\frac{Q^2K^2R^4}{\Theta^2+K^2R^4}(r'_\rho)^2\,. \tag{1-23}$$

In next section, we discuss the Reissner-Nordström metric as an example.

## 2 The case of the Reissner-Nordström metric

Similar to the Schwarzschild metric, for the Reissner-Nordström metric indicated by the following line element:

$$\mathrm{d}s^2=-\left(1-\frac{r_s}{r}+\frac{q^2}{r^2}\right)\mathrm{d}t^2+\frac{1}{1-\frac{r_s}{r}+\frac{q^2}{r^2}}\mathrm{d}r^2+r^2\left(\mathrm{d}\theta^2+\sin^2\theta\mathrm{d}\varphi^2\right), \tag{2-1}$$

we see that $\dfrac{\partial|g_{ij}|}{\partial t}=\dfrac{\partial}{\partial t}\left(\dfrac{r^4\sin^2\theta}{1-\dfrac{r_s}{r}+\dfrac{q^2}{r^2}}\right)=0$ in the area of $1-\dfrac{r_s}{r}+\dfrac{q^2}{r^2}>0$, hence, according to (0-2), in the area of $1-\dfrac{r_s}{r}+\dfrac{q^2}{r^2}>0$, the negative kinetic energy term of gravitation field vanishes.

But (2-1) cannot be continued into the area of $1-\dfrac{r_s}{r}+\dfrac{q^2}{r^2}<0$. For continuing (2-1) into the area of $1-\dfrac{r_s}{r}+\dfrac{q^2}{r^2}<0$ and, at the same time, obtaining a metric that satisfies the positive kinetic energy coordinate condition (0-5), we use the method given by §1.

For the sake of brevity, we set:

$$\frac{r}{r_s}\to r,\ \frac{t}{r_s}\to t,\ \frac{\mathrm{d}s}{r_s}\to \mathrm{d}s;\ b^2\equiv\frac{q^2}{m^2}, \tag{2-2}$$

the Reissner-Nordström line element (2-1) thus becomes

$$\mathrm{d}s^2=-\left(1-\frac{1}{r}+\frac{b^2}{4}\frac{1}{r^2}\right)\mathrm{d}t^2+\frac{1}{1-\dfrac{1}{r}+\dfrac{b^2}{4}\dfrac{1}{r^2}}\mathrm{d}r^2+r^2\left(\mathrm{d}\theta^2+\sin^2\theta\mathrm{d}\varphi^2\right). \tag{2-3}$$

Comparing (2-3) with (1-1), for the case of $1-\dfrac{1}{r}+\dfrac{b^2}{4}\dfrac{1}{r^2}>0$, we have

$$K(t,r)=\sqrt{1-\frac{1}{r}+\frac{b^2}{4}\frac{1}{r^2}},\ Q(t,r)=\frac{1}{K(t,r)}=\frac{1}{\sqrt{1-\dfrac{1}{r}+\dfrac{b^2}{4}\dfrac{1}{r^2}}},\ R(t,r)=r. \tag{2-4}$$

Notice that all the *K*, *Q* and *R* are independent of time *t*, we therefore assume that all the functions *F*, *J* and $\Theta$ in (1-16) and (1-20) are also independent of time *t*, (1-20) thus becomes

$$\frac{\mathrm{d}\Theta(r)}{\mathrm{d}r}=0,$$

we obtain immediately

$$\Theta(r)\equiv\frac{1}{A}=\text{constant}.$$

According to (1-20) and (2-4), we have

$$J(r)=\frac{\Theta(r)}{\sqrt{\Theta^2(r)+K^2(r)R^4(r)}}=\frac{1}{r^2}\frac{1}{\sqrt{f(r)}}, \tag{2-5}$$

$$f(r)=\frac{1}{r^4}+A\left(1-\frac{1}{r}+\frac{b^2}{4}\frac{1}{r^2}\right)=\frac{A}{r^4}\left(r^4-r^3+\frac{b^2}{4}r^2+\frac{1}{A}\right). \tag{2-6}$$

And, further, if we set $A>0$, then considering (2-5) we must choose the constant *A* such that $f(r)>0$ for the total space $r>0$. There are infinite such constant *A* satisfying the condition, for example,

$$\begin{aligned}&A=4,\ f(r)=1+2\left(1-\frac{1}{r}\right)^2+\left(1-\frac{1}{r^2}\right)^2+\frac{b^2}{r^2}>0;\\&A=2\left(2\pm\sqrt{2}\right),\ f(r)=2\left(1\pm\frac{1}{\sqrt{2}}-\frac{1}{r}\right)^2+\left(1-\frac{1}{r^2}\right)^2+\left(1\pm\frac{1}{\sqrt{2}}\right)\frac{b^2}{r^2}>0;\text{etc.}\end{aligned} \tag{2-7}$$

According to (1-16), (1-22) and (1-23) we now have

$$F(r)=\frac{Q(r)}{K(r)}J(r)=\frac{1}{r^2-r+\frac{b^2}{4}}\frac{1}{\sqrt{f(r)}}\,. \tag{2-8}$$

$$t'_\rho=F(r)r'_\rho=\frac{1}{\sqrt{A}}\frac{r^2}{r^2-r+\frac{b^2}{4}}\frac{1}{\sqrt{r^4-r^3+\frac{b^2}{4}r^2+\frac{1}{A}}}r'_\rho\,, \tag{2-9}$$

$$\Omega^2(\rho,\sigma)=\frac{A}{f(r)}(r'_\rho)^2>0\,. \tag{2-10}$$

From (2-9) we have

$$t(\rho,\sigma)=\int^r F(r)\mathrm{d}r+V(\sigma)=\frac{1}{\sqrt{A}}\int^r\frac{r^2}{r^2-r+\frac{b^2}{4}}\frac{\mathrm{d}r}{\sqrt{r^4-r^3+\frac{b^2}{4}r^2+\frac{1}{A}}}+V(\sigma)\,, \tag{2-11}$$

where $V(\sigma)$ is only a function of the variable $\sigma$. From the above expression we have

$$t'_\sigma=F(r)r'_\sigma+\frac{\mathrm{d}V(\sigma)}{\mathrm{d}\sigma}\,, \tag{2-12}$$

Substituting (2-12) and (2-8) to (1-4), and considering (2-4), we have

$$\frac{A}{f(r)}\left(r'_\sigma\right)^2-2\frac{1}{r^2}\frac{1}{\sqrt{f(r)}}\frac{\mathrm{d}V(\sigma)}{\mathrm{d}\sigma}r'_\sigma+W^2-K^2(r)\left(\frac{\mathrm{d}V(\sigma)}{\mathrm{d}\sigma}\right)^2=0\,; \tag{2-13}$$

If we take

$$W(\rho,\sigma)=\frac{\mathrm{d}V(\sigma)}{\mathrm{d}\sigma}W_0(\rho,\sigma)\,, \tag{2-14}$$

then from (2-13) we obtain

$$r'_\sigma=\frac{\mathrm{d}V(\sigma)}{\mathrm{d}\sigma}\frac{\sqrt{f(r)}}{A}\left[\frac{1}{r^2}+\delta\sqrt{f(r)-AW_0^2}\right], \tag{2-15}$$

where $\delta=\pm1$; and we must choose $W_0(\rho,\sigma)$ that satisfies $f(r)-AW_0^2\geq 0$. For example, if $A=4$ then we can choose that $W_0(\rho,\sigma)=\frac{1}{2}$, or $W_0(\rho,\sigma)=\frac{1}{2}\sqrt{1+2\left(1-\frac{1}{r}\right)^2}$, etc; or we choose that $W_0(\rho,\sigma)=\frac{1}{\sqrt{A}}\sqrt{f(r)}$ for arbitrary constant *A*.

From (2-15) we obtain

$$\int^r\frac{A}{\sqrt{f(r)}}\frac{\mathrm{d}r}{\frac{1}{r^2}+\delta\sqrt{f(r)-AW_0^2}}=\int^r\frac{Ar^2}{\sqrt{f(r)}}\frac{\mathrm{d}r}{1+\delta\, r^2\sqrt{f(r)-AW_0^2}}=U(\rho)+V(\sigma)\,. \tag{2-16}$$

where $U(\rho)$ is only a function of the variable $\rho$. Hence, after designating the functions $W_0(\rho,\sigma)$, $U(\rho)$ and $V(\sigma)$, from (2-16) and (2-11) we obtain the transformational relations $t=t(\rho,\sigma)$, $r=r(\rho,\sigma)$ between $(t,r)$ and $(\rho,\sigma)$.

From (2-16) we have

$$r'_\rho=\frac{\mathrm{d}U(\rho)}{\mathrm{d}\rho}\frac{\sqrt{f(r)}}{A}\left[\frac{1}{r^2}+\delta\sqrt{f-AW_0^2}\right]. \tag{2-17}$$

Substituting (2-17) to (2-10), we obtain

$$\Omega(\rho,\sigma)=\frac{1}{\sqrt{A}}\frac{1+\delta\, r^2\sqrt{f(r)-AW_0^2}}{r^2}\frac{\mathrm{d}U(\rho)}{\mathrm{d}\rho}\,; \tag{2-18}$$

And, further, substituting (2-9), (2-12), (2-17), (2-15) and (2-4) to (1-6), we obtain

$$E(\rho,\sigma)=\frac{1}{A}\delta\sqrt{f(r)-AW_0^2}\left[\frac{1}{r^2}+\delta\sqrt{f(r)-AW_0^2}\right]\frac{\mathrm{d}U(\rho)}{\mathrm{d}\rho}\frac{\mathrm{d}V(\sigma)}{\mathrm{d}\sigma}. \quad (2\text{-}19)$$

If we choose appropriate forms of the functions $W_0(\rho,\sigma)$, $U(\rho)$ and $V(\sigma)$, then we can obtain simpler forms of $W(\rho,\sigma)$, $\Omega(\rho,\sigma)$ and $E(\rho,\sigma)$. For example, when $A=4$, if we designate $W_0(\rho,\sigma)=\frac{1}{\sqrt{2}}$, $V(\sigma)=\sqrt{2}\sigma$, then from (2-14) we see $W(\rho,\sigma)=1$.

If we take

$$W_0(\rho,\sigma)=\frac{1}{\sqrt{A}}\sqrt{f(r)}, \quad (2\text{-}20)$$

then from (2-19) we see $E(\rho,\sigma)=0$. And, further, we designate that

$$U(\rho)=\sqrt{A}\rho, V(\sigma)=\pm\sqrt{A}\sigma, \quad (2\text{-}21)$$

according to (2-14), (2-18) and (2-19), the line element (1-3) in this case becomes

$$\mathrm{d}s^2=-\left[\frac{r_s^4}{r^4}+A\left(1-\frac{r_s}{r}+\frac{q^2}{r^2}\right)\right]\mathrm{d}\sigma^2+\frac{r_s^4}{r^4}\mathrm{d}\rho^2+r^2(\mathrm{d}\theta^2+\sin^2\theta\mathrm{d}\varphi^2); \quad (2\text{-}22)$$

we see that the unique singularity in (2-22) appears at $r = 0$.

(2-16) and (2-11) now become

$$\int^r\frac{r^4}{r_s^4}\frac{\mathrm{d}r}{\sqrt{\frac{r^4}{r_s^4}-\frac{r^3}{r_s^3}+\frac{b^2}{4}\frac{r^2}{r_s^2}+\frac{1}{A}}}=\rho\pm\sigma, \quad (2\text{-}23)$$

$$t(\rho,\sigma)=\frac{1}{\sqrt{A}}\int^r\frac{r^2}{r_s^2}\frac{1}{\frac{r^2}{r_s^2}-\frac{r}{r_s}+\frac{b^2}{4}}\frac{\mathrm{d}r}{\sqrt{\frac{r^4}{r_s^4}-\frac{r^3}{r_s^3}+\frac{b^2}{4}\frac{r^2}{r_s^2}+\frac{1}{A}}}\pm\sqrt{A}\sigma. \quad (2\text{-}24)$$

In (2-22), (2-23) and (2-24), (2-2) is considered.

Similar to the case of the Lemaitre metric of the Schwarzschild solution, (2-22), (2-23) and (2-24) show that both the metrics determined by $r=r(\rho,\sigma)$ and its time-reverse $r=r(\rho,-\sigma)$ respectively are the solution of the Einstein equation $R^{\mu\nu}-\frac{1}{2}g^{\mu\nu}R=\frac{8\pi G}{c^4}T^{\mu\nu}_{(\mathrm{EM})}$, where $T^{\mu\nu}_{(\mathrm{EM})}$ is the energy-momentum tensor of the electromagnetic field of a point charge with the charge $q$.

Generally speaking, both the integrals in (2-23) and (2-24) have to be expressed by the elliptical integral, respectively. For example, if we use one of two values of $A$ given by (2-7), then we must take advantage of the elliptical integral for both the integrals in (2-23) and (2-24). On the other hand, if $q^2\ge m^2$, then $1-\frac{r_s}{r}+\frac{q^2}{r^2}=\left(1-\frac{m}{r}\right)^2+\frac{q^2-m^2}{r^2}\ge 0$. We therefore can always use the original form (2-1) of the Reissner-Nordström metric but not need to make a transformation (1-2) in this case (of course, there is still a coordinate singularity $r=m$ in the original form (2-1) when $q^2=m^2$), what conclusions we have obtained in this section are only needed for the case of $b^2=\frac{q^2}{m^2}<1$. For this case, we set

$$\frac{r_0}{r_s}=\frac{3+\sqrt{9-8b^2}}{8}>0, \quad (2\text{-}25)$$

$$\omega^2\equiv\frac{r_0}{r_s}-\frac{1}{2}=\frac{\sqrt{9-8b^2}-1}{8}>0; \quad (2\text{-}26)$$

We can prove that the quartic equation $\frac{r^4}{r_s^4}-\frac{r^3}{r_s^3}+\frac{b^2}{4}\frac{r^2}{r_s^2}+\frac{1}{A}=0$ has repeated root if and only if

$$A=\frac{1}{\omega^2}\left(\frac{r_s}{r_0}\right)^3>0, \tag{2-27}$$

and, further, if we take (2-27), then we have

$$\frac{r^4}{r_s^4}-\frac{r^3}{r_s^3}+\frac{b^2}{4}\frac{r^2}{r_s^2}+\frac{1}{A}=\frac{r^4}{r_s^4}-\frac{r^3}{r_s^3}+\frac{b^2}{4}\frac{r^2}{r_s^2}+\omega^2\left(\frac{r_0}{r_s}\right)^3=\left(\frac{r}{r_s}-\frac{r_0}{r_s}\right)^2\left[\left(\frac{r}{r_s}+\omega^2\right)^2+\frac{1}{2}\omega^2\right], \tag{2-28}$$

And from (2-6) we have

$$f(r)=\frac{1}{\omega^2}\left(\frac{r_s}{r_0}\right)^3\frac{r_s^4}{r^4}\left(\frac{r}{r_s}-\frac{r_0}{r_s}\right)^2\left[\left(\frac{r}{r_s}+\omega^2\right)^2+\frac{1}{2}\omega^2\right]\geq 0,$$

we see that (2-27) is allowable.

According to (2-28), (2-22), (2-23) and (2-24) now become

$$\mathrm{d}s^2=-\frac{1}{\omega^2}\left(\frac{r_s}{r_0}\right)^3\frac{r_s^4}{r^4}\left(\frac{r}{r_s}-\frac{r_0}{r_s}\right)^2\left[\left(\frac{r}{r_s}+\omega^2\right)^2+\frac{1}{2}\omega^2\right]\mathrm{d}\sigma^2+\frac{r_s^4}{r^4}\mathrm{d}\rho^2+r^2(\mathrm{d}\theta^2+\sin^2\theta\mathrm{d}\varphi^2), \tag{2-29}$$

$$\int^r\frac{r^4}{r_s^4}\frac{1}{\frac{r}{r_s}-\frac{r_0}{r_s}}\frac{\mathrm{d}r}{\sqrt{\left(\frac{r}{r_s}+\omega^2\right)^2+\frac{1}{2}\omega^2}}=\rho\pm\sigma, \tag{2-30}$$

$$t(\rho,\sigma)=\omega\sqrt{\left(\frac{r_0}{r_s}\right)^3}\int^r\frac{r^2}{r_s^2}\frac{1}{\frac{r}{r_s}-\frac{r_0}{r_s}}\frac{1}{\frac{r^2}{r_s^2}-\frac{r}{r_s}+\frac{b^2}{4}}\frac{\mathrm{d}r}{\sqrt{\left(\frac{r}{r_s}+\omega^2\right)^2+\frac{1}{2}\omega^2}}\pm\frac{1}{\omega}\sqrt{\left(\frac{r_s}{r_0}\right)^3}\sigma. \tag{2-31}$$

We see that both the integrals in (2-30) and (2-31) can now be expressed by elementary functions. However, a weak point of (2-29) is that the coefficient of $\mathrm{d}\sigma^2$ vanishes when $r=r_0$.

If the charge $q=0$, then what metric that satisfies the positive kinetic energy coordinate condition (0-5) we obtain in this section corresponds to the case of the Schwarzschild solution.

## 3 The case of the Tolman metric

We now discuss simply a spherically symmetric co-moving coordinate system, which plays a basic part in some researches, for example, gravitational collapse and the Robertson-Walker metric[4].

### 3.1 The Tolman metric

The metric of a spherically symmetric co-moving coordinate system are indicated by the following line element:

$$\mathrm{d}s^2=-\mathrm{d}t^2+\mathrm{e}^{\alpha(t,r)}\mathrm{d}r^2+\beta^2(t,r)(\mathrm{d}\theta^2+\sin^2\theta\mathrm{d}\varphi^2); \tag{3-1}$$

We take the energy-momentum tensor of matter

$$T^{\mu\nu}=pg^{\mu\nu}+(\rho+p)U^\mu U^\nu, \tag{3-2}$$

where

$$\rho=\rho(t,r),\ p=p(t,r);U^\mu=(1,0,0,0). \tag{3-3}$$

The Einstein equation $R_{\mu\nu}=8\pi G\left(T_{\mu\nu}-\frac{1}{2}g_{\mu\nu}T\right)$ leads to the following equations:

$$R_{00}=-\frac{\ddot{\alpha}}{2}-\frac{\dot{\alpha}^2}{4}-2\frac{\ddot{\beta}}{\beta}=4\pi G(\rho+3p),\tag{3-4}$$

$$R_{01}=-2\frac{\dot{\beta}'}{\beta}+\frac{\dot{\alpha}\beta'}{\beta}=0\ ,\tag{3-5}$$

$$R_{11}=\mathrm{e}^{\alpha}\left(\frac{\ddot{\alpha}}{2}+\frac{\dot{\alpha}^2}{4}+\frac{\dot{\alpha}\dot{\beta}}{\beta}\right)-2\frac{\beta''}{\beta}+\frac{\alpha'\beta'}{\beta}=4\pi G\mathrm{e}^{\alpha}(\rho-p),\tag{3-6}$$

$$R_{22}=\mathrm{e}^{-\alpha}\left(\frac{\alpha'\beta\beta'}{2}-\beta'^2-\beta\beta''\right)+\beta\ddot{\beta}+\dot{\beta}^2+\frac{\dot{\alpha}\beta\dot{\beta}}{2}+1=4\pi G\beta^2(\rho-p),\tag{3-7}$$

$$R_{33}=R_{22}\sin^2\theta=4\pi G\beta^2\sin^2\theta(\rho-p).\tag{3-8}$$

Form (3-5) we have

$$\dot{\alpha}=2\frac{\dot{\beta}'}{\beta'}=\frac{\partial}{\partial t}\left[\ln\left(\beta'^2\right)\right],\tag{3-9}$$

we therefore obtain

$$\mathrm{e}^{\alpha}=\frac{\beta'^2}{1+f(r)},f(r)>-1.\tag{3-10}$$

The form of (3-1) with (3-10) is so called the Tolman metric.

From (3-5) we have $\ddot{\alpha}=2\frac{\ddot{\beta}'}{\beta'}-2\left(\frac{\dot{\beta}'}{\beta'}\right)^2$, substituting this formula and (3-9) to (3-4), we obtain

$$\frac{\ddot{\beta}'}{\beta'}+2\frac{\ddot{\beta}}{\beta}=-4\pi G(\rho+3p).\tag{3-11}$$

Calculating $(3\text{-}7)-\frac{1}{2}\beta^2\left((3\text{-}5)+\mathrm{e}^{-\alpha}(3\text{-}6)\right)$ and using (3-10) we have

$$2\beta\ddot{\beta}+\dot{\beta}^2-f(r)=-8\pi G\beta^2p\ ,\tag{3-12}$$

from the above formula we obtain

$$\beta\left(\dot{\beta}^2-f(r)\right)=-\frac{8}{3}\pi G\int^t p\frac{\partial\beta^3}{\partial t}\mathrm{d}t+F(r)\ .\tag{3-13}$$

On the other hand, we can prove that we can only obtain three independent equations (3-10), (3-11) and (3-13) from (3-4)~(3-8).

The equation of conservation of the energy-momentum tensor $T^{\mu\nu}{}_{;\nu}=0$ can only provide two independent equations:

$$T^{0\nu}{}_{;\nu}=\frac{1}{\left(\beta^3\right)'}\left\{\frac{\partial\left[\rho\left(\beta^3\right)'\right]}{\partial t}+p\frac{\partial\left[\left(\beta^3\right)'\right]}{\partial t}\right\}=0,\tag{3-14}$$

$$T^{1\nu}{}_{;\nu}=\mathrm{e}^{-\alpha}\frac{\partial p}{\partial r}=0\ .\tag{3-15}$$

From (3-15) we obtain

$$p=p(t)\ ;\tag{3-16}$$

from (3-11), (3-13) and (3-15) we obtain

$$\rho=-\frac{1}{\left(\beta^3\right)'}\int^t p(t)\frac{\partial\left[\left(\beta^3\right)'\right]}{\partial t}\mathrm{d}t+\frac{3}{8\pi G}\frac{F'(r)}{\left(\beta^3\right)'}.\tag{3-17}$$

We can prove that we can only obtain four independent equations (3-10), (3-13), (3-16) and (3-17) from (3-4)~(3-8), (3-14) and (3-15), notice that (3-11) now is no longer an independent equation.

Of course, we should have an equation of state

$$p = p(\rho)\,. \tag{3-18}$$

If we try to seek a separable solution of $\beta$ :

$$\beta = \beta_0(t)\beta_1(r)\,, \tag{3-19}$$

then (3-13) becomes

$$\beta_0(t)\left(\dot{\beta}_0^2(t) - \frac{f(r)}{\beta_1^2(r)}\right) = -\frac{8}{3}\pi G\int^t p(t)\frac{\partial\beta_0^3(t)}{\partial t}\mathrm{d}t + \frac{F(r)}{\beta_1^3(r)}\,,$$

we see that, for this case, we must take

$$f(r) = -k\beta_1^2(r)\,, F(r) = k'\beta_1^3(r)\,, \tag{3-20}$$

where both $k$ and $k'$ are constants. And $\beta_0(t)$ satisfies

$$\beta_0(t)\left(\dot{\beta}_0^2(t) + k\right) = -\frac{8}{3}\pi G\int^t p(t)\frac{\partial\beta_0^3(t)}{\partial t}\mathrm{d}t + k'\,. \tag{3-21}$$

Substituting (3-19) and (3-20) to (3-17), we obtain

$$\rho = -\frac{1}{\beta_0^3(t)}\int^t p(t)\frac{\partial\left(\beta_0^3(t)\right)}{\partial t}\mathrm{d}t + \frac{3k'}{8\pi G}\frac{1}{\beta_0^3(t)} \equiv \rho(t)\,. \tag{3-22}$$

We see that, for this case, $\rho$ is consequentially only a function of $t$, namely, it is independent of $r$.

**3.2 The Tolman metric that satisfies the positive kinetic energy coordinate condition is inexistent**

If we ask that the metric indicated by (3-1) and (3-10) satisfies the positive kinetic energy coordinate condition (0-5), then we have

$$\left(\sqrt{|g_{ij}|}\frac{g^{0\lambda}}{g^{00}}\right)_{,\lambda} = \sin\theta\frac{\partial\left(\mathrm{e}^{\alpha}\beta^2\right)}{\partial t} = \sin\theta\frac{\partial}{\partial t}\left(\frac{\beta^2\beta'}{\sqrt{1+f(r)}}\right)$$

$$= \frac{1}{3}\frac{\sin\theta}{\sqrt{1+f(r)}}\frac{\partial\left[\left(\beta^3\right)'\right]}{\partial t} = \frac{1}{3}\frac{\sin\theta}{\sqrt{1+f(r)}}\frac{\partial^2\left(\beta^3\right)}{\partial t\partial r} = 0\,.$$

The solution of the above equation reads:

$$\beta = \sqrt[3]{\tilde{\beta}(t) + \bar{\beta}(r)}\,. \tag{3-23}$$

Substituting (3-23) to (3-17), we obtain

$$\rho = \frac{3}{8\pi G}\frac{F'(r)}{\bar{\beta}'(r)} \equiv \rho(r)\,. \tag{3-24}$$

We see that, for this case, $\rho$ is consequentially only a function of $r$, namely, it is independent of $t$.

From (3-23) we have $\frac{\ddot{\beta}'}{\beta'} + 2\frac{\ddot{\beta}}{\beta} = \frac{2}{3}\left(\frac{\dot{\tilde{\beta}}(t)}{\tilde{\beta}(t)+\bar{\beta}(r)}\right)^2$. Substituting this formula, (3-16) and (3-24) to (3-11), we obtain

$$\frac{2}{3}\left(\frac{\dot{\tilde{\beta}}(t)}{\tilde{\beta}(t)+\bar{\beta}(r)}\right)^2 = -4\pi G\left(\rho(r) + 3p(t)\right)\,. \tag{3-25}$$

It is obvious that (3-25) has not solution for general case. Hence, the Tolman metric that satisfies the positive kinetic energy coordinate condition is inexistent.

### 3.3 The coordinate transformation between the Tolman metric and a metric satisfying the positive kinetic energy coordinate condition

Now that the Tolman metric that satisfies the positive kinetic energy coordinate condition is inexistent, we have to use the method given by §1, namely, we make the coordinate transformation (1-2) for seeking a coordinate system $(\sigma, \rho, \theta, \varphi)$ satisfying the positive kinetic energy coordinate condition for the given spherically symmetric co-moving coordinate system $(t, r, \theta, \varphi)$. According to the discussion in §1, we must solve the equation (1-21) as the first step for this purpose. Substituting the Tolman metric indicated (3-1) and (3-10) to (1-21), we have

$$\frac{\partial \Theta}{\partial r} + \frac{1}{\sqrt{1+f(r)}} \frac{\partial}{\partial t}\left( \beta' \sqrt{\Theta^2 + \beta^4} \right) = 0 . \qquad (3\text{-}26)$$

Especially, for the separable solution (3-19) of $\beta$, if we redefine the radial coordinate such that $\beta_1(r) = r$, then according to (3-20), (3-26) becomes

$$\frac{\partial \Theta}{\partial r} + \frac{1}{\sqrt{1-kr^2}} \frac{\partial}{\partial t}\left( \beta_0(t) \sqrt{\Theta^2 + r^4 \beta_0^4(t)} \right) = 0 . \qquad (3\text{-}27)$$

Once we obtain the relation between $(\sigma, \rho, \theta, \varphi)$ and $(t, r, \theta, \varphi)$, the results obtained in the spherically symmetric co-moving coordinate system $(t, r, \theta, \varphi)$ should be explained in the coordinate system $(\sigma, \rho, \theta, \varphi)$.